\documentclass[12pt]{article}
\usepackage{geometry,amsmath,amssymb,enumerate,epsfig}
\newcommand{\Chi}{\mathrm{X}}
\newcommand{\Iota}{\mathrm{I}}
\newcommand{\Omicron}{\mathrm{O}}
\newcommand{\assign}{:=}
\newcommand{\emdash}{---}
\newcommand{\mathe}{\mathrm{e}}
\newcommand{\tmem}[1]{{\em #1\/}}

\newcommand{\tmop}[1]{\ensuremath{\operatorname{#1}}}
\newcommand{\tmtextbf}[1]{{\bfseries{#1}}}

\newcommand{\tmtexttt}[1]{{\ttfamily{#1}}}
\newenvironment{enumeratealpha}{\begin{enumerate}[a{\textup{)}}] }{\end{enumerate}}
\newenvironment{enumerateroman}{\begin{enumerate}[i.] }{\end{enumerate}}
\newenvironment{enumerateromancap}{\begin{enumerate}[I.] }{\end{enumerate}}

\begin{document}

\title{Simulating the All-Order Strong Coupling Expansion III:
O($N $) sigma/loop models
}
\author{Ulli Wolff\thanks{
e-mail: uwolff@physik.hu-berlin.de} \\
Institut f\"ur Physik, Humboldt Universit\"at\\ 
Newtonstr. 15 \\ 
12489 Berlin, Germany
}
\date{}
\maketitle

\begin{abstract}
  We reformulate the O($N$) sigma model as a loop model whose configurations
  are the all-order strong coupling graphs of the original model. The loop
  configurations are represented by a pointer list in the computer and a Monte
  Carlo update scheme is proposed. Sample simulations are reported and the
  method turns out to be similarly efficient as the reflection cluster method,
  but it has greater potential for systematic generalization to other lattice
  field theories. A variant action suggested by the method is also simulated
  and leads to a rather extreme demonstration of the concept of universality
  of the scaling or continuum limit.

\end{abstract}
\begin{flushright} HU-EP-09/34 \end{flushright}
\begin{flushright} SFB/CCP-09-69 \end{flushright}
\thispagestyle{empty}
\newpage

\section{Introduction}

The family of globally O($N$) invariant nonlinear sigma models, also called
$N$-vector models, are very important statistical systems. For obvious
reasons, in three space dimensions, they play a very prominent r\^ole in
condensed matter physics. We here only mention the XY model ($N = 2$),
relevant for the description of liquid helium, and the Heisenberg model ($N =
3$) for magnets. In high energy physics the four dimensional versions appear
as effective field theories, for instance for pion physics, and a lot of
interest focuses also on two (Euclidean) dimensions. This is motivated by the
fact that these field theories are asymptotically free and share features with
QCD like asymptotic freedom and dimensional transmutation with the
nonperturbative generation of a scale like $\Lambda_{\tmop{QCD}}$. For example
the study of a nonperturbative renormalized coupling constant in
{\cite{Luscher:1991wu}} was an important preparation for the QCD Schr\"odinger
functional methods {\cite{Luscher:1992an}}.

With regard to the technique of Monte Carlo simulation, our main access to the
models beyond perturbation theory, since about 20 years we are in an
exceptional situation with regard to the O($N$) models. The method of cluster
updates {\cite{Swendsen:1987ce}} for embedded Z(2) (Ising) degrees of freedom
{\cite{Wolff:1988uh}} allows to painlessly enter the critical region, which
unfortunately is in stark contrast to our possibilities in QCD. The hope that
the cluster method would be widely generalizable was unfortunately
disappointed in the following years, at least for high energy physics. In
{\cite{Caracciolo:1992nh}} even a kind of `heuristic{\footnote{There are
mathematical proofs which refer however to smooth field configurations.}}
no-go theorem' was given concerning the generalization to sigma models with
other spin manifolds. With the advent (or rather recognition{\footnote{I am
indebted to Urs Wenger in this context.}}) of {\cite{prokofev2001wacci}} a
completely different strategy to overcome slowing down has appeared: Simulate
the strong coupling graphs (to arbitrary order) instead of field
configurations. Close to criticality the relevant field configurations are
long-distance correlated. The cluster method, using an auxiliary percolation
process, manages to execute collective moves that are thermodynamically
appropriate for this case. Such moves seem to be difficult to find (and
implement efficiently) in general{\footnote{See {\cite{Luscher:2009eq}} for a
recent new proposal in this direction.}}. The strong coupling graphs that are
relevant at criticality are large and numerous. The clever idea of Prokof'ev
and Svistunov {\cite{prokofev2001wacci}} was to generate them not for the
partition function alone, but to simultaneously consider the two point
correlation. They have demonstrated in simple models that local deformations
can pass between such graphs without significant obstructions and thus a
relevant sample can be simulated. To avoid confusion we remind the reader of
the following. In ordinary strong coupling expansions one takes the
thermodynamic limit term by term and then the series usually has a finite
radius of convergence which is often related to phase transitions. A finite
lattice regularizes such singularities (for compact fields at least) and we
can compute everything to in principle arbitrary precision by a convergent
expansion for arbitrary couplings or temperature. While close to criticality
this is impractical by conventional systematic expansion the stochastic
evaluation is feasible. We may now also speak about an equivalence with
another statistical system which incidentally has only discrete variables.
Note that, although there are some similarities, this is not a complete
duality transformation in the sense of Kramers and Wannier
{\cite{PhysRev.60.252}}.

In a recently begun series of papers {\cite{Wolff:2008km}},
{\cite{Wolff:2008xa}} we have started to further work out the new approach.
Beyond the Ising model we could apply it to fermions. Due to the sign problem
this is at the moment still restricted to two dimensional systems like the
Gross-Neveu model {\cite{Bar:2009yq}}. A novelty that one has to appreciate is
that the generated graphs can be adapted to the observables that one is
interested in. One then needs several simulations for different quantities. We
nonetheless see these dedicated simulations as a strength of the method. In
{\cite{Wolff:2008km}}, {\cite{Wolff:2009ke}}, {\cite{Bar:2009yq}} it was found
that this extra effort can result in enhanced precision for interesting
observables.

In this article we successfully extend the all-order strong coupling method to
the class of O($N$) nonlinear sigma models in arbitrary dimension. We first
achieve this for the standard lattice action which allows to confirm our
results by comparing to other data in the literature. While the standard
action has a relatively complicated all-order expansion we can define another
action by insisting on a simpler expansion. Formally it can be argued to lie
in the same universality class, but on the other hand it looks like a rather
radical mutilation of the original spin model. We simulate its graphs and find
that at least one universal result is reproduced quite accurately and
universality is confirmed. This flexibility will hopefully be useful to tackle
further more complicated models in the future.

In section 2 we develop the loop model equivalent to the all-order expansion
of the O($N$) system. In section 3 we discuss how to represent the loop
configurations in the computer and how to sample them. Extensive tests with
the standard action are carried out in section 4. Section 5 discusses
universality and the modified action followed by conclusions in section 6. In an
appendix the limit $N = 1$ of our algorithm and its relation to previous Ising
work is discussed.

\section{O($N$) model as a loop ensemble}

We consider spin models with $N$ component spins $s (x)$ of unit length
located at the sites of a $D$ dimensional hypertorus of length $L_{\mu}$ in
the various directions. We refer all lengths to the isotropic lattice spacing
thus putting $a = 1$. For the standard lattice action the partition function
with two field insertions reads
\begin{equation}
  Z (u, v) = \int \left[ \prod_z d \mu (s (z)) \right] \mathe^{\beta \sum_{l =
  \langle x y \rangle} s (x) \cdot s (y)} s (u) \cdot s (v) . \label{Zuv}
\end{equation}
The sum is over nearest neighbor links and the dots between pairs of spins
mean O($N$) invariant scalar products. The integrations employ the normalized
O($N$) invariant measure on the sphere,
\begin{equation}
  \int d \mu (s) f (s) = K_N \int d^N s \delta (s^2 - 1) f (s), \hspace{1em}
  K_N \leftrightarrow \int d \mu (s) = 1.
\end{equation}
Later on we shall need the corresponding single site generating function for a
general source $j_{\alpha}$
\begin{equation}
  \int d \mu (s) \mathe^{j \cdot s} = G_N (j) = \sum_{n = 0}^{\infty} c [n ;
  N] (j \cdot j)^n \label{Gdef}
\end{equation}
which is essentially given by the modified Bessel function $I_{N / 2 - 1}$ and
has expansion coefficients
\begin{equation}
  c [n ; N] = \frac{\Gamma (N / 2)}{2^{2 n} n! \Gamma (N / 2 + n)} .
\end{equation}

The strong coupling expansion in $\beta$ is generated by independently summing
over an integer link field $k (l) = 0, 1, 2, \ldots, \infty$ in
\begin{equation}
  Z (u, v) = \sum_k \int \left[ \prod_z d \mu (s (z)) \right] \prod_{l =
  \langle x y \rangle} \frac{\beta^{k (l)}}{k (l) !} [s (x) \cdot s (y)]^{k
  (l)} s (u) \cdot s (v) . \label{Zsuv}
\end{equation}
For a given configuration $k$ the spin integral may now be written as
\begin{equation}
  X = \frac{\partial}{\partial j_{\alpha} (u)}  \frac{\partial}{\partial
  j_{\alpha} (v)} \prod_{l = \langle x y \rangle} \left[
  \frac{\partial}{\partial j_{\gamma} (x)}  \frac{\partial}{\partial
  j_{\gamma} (y)} \right]^{k (l)} \prod_z G_N (j (z)) |_{j \equiv 0} .
\end{equation}
We next introduce an auxiliary integer site field
\begin{equation}
  d (x) = \delta_{x, u} + \delta_{x, v} + \sum_{l, \partial l \ni x} k (l)
  \label{ddef}
\end{equation}
which counts the number of spins or respectively $j$-derivatives at $x$. It as
well as $X$ depends on $u, v, k$, of course, which we leave implicit for
easier notation. To produce a nonzero $X, d (x)$ has to be even on all sites.
Then the contribution becomes
\begin{equation}
  X' = \frac{\partial}{\partial j_{\alpha} (u)}  \frac{\partial}{\partial
  j_{\alpha} (v)} \prod_{l = \langle x y \rangle} \left[
  \frac{\partial}{\partial j_{\gamma} (x)}  \frac{\partial}{\partial
  j_{\gamma} (y)} \right]^{k (l)} \prod_z [j (z) \cdot j (z)]^{d (z) / 2}
\end{equation}
where there are as many $j$ factors as there are derivatives. In addition $X'$
differs from $X$ by dropping factors $c [d (z) / 2 ; N]$ for all sites $z$.
The total number of terms in $X'$ from taking all derivatives is
\begin{equation}
  \mathcal{M}_0 [u, v ; k] = \prod_z d (z) ! .
\end{equation}
The terms differ in their O($N$) index contraction structures. To each of them
there corresponds a graph $\Lambda$ drawn on the lattice. There are $k
(l)$ lines between each nearest neighbor pair $\langle x y \rangle = l$. At
the `interior of the sites' there is a kind of switch-board that sets up
pairwise connections between all surrounding lines. Only at $u$ and $v$ two
lines are left unpaired locally and instead are contracted with each other.
Thus all lines are arranged in closed loops. The chain between $u$ and $v$
does not close geometrically (unless $u = v$) but closes with respect to
O($N)$ contractions leading to a factor $N$ as all other loops do. In the next
section a visualization of such a graph or loop configuration will be given.
Each graph represents a subset of the $\mathcal{M}_0$ terms. This multiplicity is
given by
\begin{equation}
  \mathcal{M}[\Lambda] = \frac1{\mathcal{S}[\Lambda]}
  \left( \prod_l k (l) ! \right) \prod_x [d (x) / 2] !
  2^{d (x) / 2}.
\end{equation}
The last two factors correspond to permuting 
the pairs $j \cdot j$ at the sites and the two factors in each pair.
In addition we consider the permutation of lines on the same link.
For some graphs this leads however to an overcounting which is canceled
by the symmetry factor $ \mathcal{S}[\Lambda]$. It is given by the
number of elements of the group of line permutations which leave
the connectivity of the graph unchanged. Apart from dealing with graphs
embedded on a lattice it is
similar to the symmetry factors that also appear for Feynman diagrams.
We return to this issue
in the discussion of our update scheme for graphs $\Lambda$.
So far we have considered the loop
configurations for given $u, v, k$. It is clear however, that the latter are
also determined by the graph on the lattice. We may hence independently sum
over graphs $\Lambda \in \mathcal{L}_2$ which we define to include all
possible locations $u, v$ of the two `defects' and all possible $k (l)$
assignments to links that produce nonvanishing contributions. Then from
(\ref{Zsuv}) we generalize to
\begin{equation}
  \mathcal{Z}= \sum_{u, v} \rho^{- 1} (u - v) Z (u, v) = \sum_{\Lambda \in
  \mathcal{L}_2} \rho^{- 1} (u - v) W [\Lambda] \label{ZLam}
\end{equation}
where we have collected the whole loop weight into
\begin{eqnarray}
  W [\Lambda] & = & N^{| \Lambda |} \mathcal{M}[\Lambda] \left[ \prod_{l =
  \langle x y \rangle} \frac{\beta^{k (l)}}{k (l) !} \right] \prod_z c [d (z)
  / 2 ; N] \nonumber\\
  & = & N^{| \Lambda |} \beta^{\sum_l k (l)} \frac1{\mathcal{S}[\Lambda]}
  \left[ \prod_x 2^{- d (x) / 2}
  \frac{\Gamma (N / 2)}{\Gamma (N / 2 + d (x) / 2)} \right] .  \label{Wdef}
\end{eqnarray}
Here \ $u, v$, $k (l)$ and thus $d (x)$ are now functions of $\Lambda$, and
$\rho$ is a positive weight to be chosen later. In the exponent $| \Lambda |$
means the number of individual closed loops including the $u$-$v$ chain. The
loops in the configurations $\Lambda$ that we sum over can overlap, intersect
and backtrack. The weight depends on these features, the loops interact. In
the next section we shall introduce an explicit parameterization of $\Lambda$
together with an update scheme to simulate the loop model.

As a by now standard next step {\cite{Wolff:2008km}} we introduce expectation
values with respect to the new ensemble
\begin{equation}
  \langle \langle A (\Lambda) \rangle \rangle = \frac{1}{\mathcal{Z}}
  \sum_{\Lambda \in \mathcal{L}_2} \rho^{- 1} (u - v) W [\Lambda] A (\Lambda)
  . \label{LGav}
\end{equation}
For the two point function of the original model there is the trivial relation
\begin{equation}
  \langle s_{\alpha} (u) s_{\beta} (v) \rangle = \frac{\delta_{\alpha
  \beta}}{N}  \frac{Z (u, v)}{Z (\emptyset)}
\end{equation}
where $Z (\emptyset)$ is the partition function without insertions (or $u =
v$). This ratio can obviously be obtained from (\ref{LGav}) as
\begin{equation}
  \langle s_{\alpha} (0) s_{\beta} (x) \rangle = \frac{\delta_{\alpha
  \beta}}{N} \rho (x) \frac{\langle \langle \delta_{u - v, x} \rangle
  \rangle}{\langle \langle \delta_{u, v} \rangle \rangle} \label{sscor}
\end{equation}
where we have assumed the normalization
\begin{equation}
  \rho (0) = 1.
\end{equation}
It is convenient to in addition introduce expectation values referring to the
subset of `vacuum' configurations $\Lambda \in \mathcal{L}_2$ that have $u =
v$,
\begin{equation}
  \langle \langle A (\Lambda) \rangle \rangle_0 = \frac{\langle \langle
  \delta_{u, v} A \rangle \rangle}{\langle \langle \delta_{u, v} \rangle
  \rangle} .
\end{equation}
Such expectation values are independent of the choice of $\rho$.

The internal energy density is equivalent to the average of the nearest
neighbor correlation
\begin{equation}
  E = \frac{1}{N_l}  \sum_{l = \langle x y \rangle} \langle s (x) \cdot s (y)
  \rangle \leqslant 1 \label{Edef}
\end{equation}
where $N_l$ is the total number of links. By differentiating $Z
(\emptyset)$=$\mathcal{Z} \langle \langle \delta_{u, v} \rangle \rangle$ in
both representations with respect to $\beta$, the relation
\begin{equation}
  \beta E = \frac{1}{N_l} \sum_l \langle \langle k (l) \rangle \rangle_0
  \assign K. \label{kobs}
\end{equation}
follows easily. Thus the average link occupation is bounded by $\beta$.
Numerical experience and large $N$ considerations have shown that deep in the
critical regime $\beta / N$ are typically numbers smaller than one. Thus,
although the total length of the loops in $\Lambda$ is unbounded in principle,
as an extensive quantity it will in practice never exceed the number of links
on the lattice (times $N$) by a large factor. We will come back to this point
when we simulate. Another standard observable is the susceptibility
\begin{equation}
  \chi = \sum_x \langle s (0) \cdot s (x) \rangle = \frac{\langle \langle \rho
  (u - v) \rangle \rangle}{\langle \langle \delta_{u, v} \rangle \rangle}
  \label{chiobs}
\end{equation}
which follows from contracting and summing over $x$ in (\ref{sscor}).

The representation derived here allows to actually set $N$ also to non-integer
values. The value $N = 1$ corresponds to the \ Ising model. The limit $N
\rightarrow 1$ of the present method does not immediately coincide with
{\cite{prokofev2001wacci}}, {\cite{Wolff:2008km}}. Therefore in appendix A we
discuss the connection.

\section{Parameterizing and simulating the loop ensemble}

\subsection{Parameterization: loops as lists}

It is often fruitful to first analyze graphs in an abstract manner and to
separately consider their embedding on the lattice {\cite{citeulike:4800116}}.
At the abstract level each graph $\Lambda \in \mathcal{L}_2$ consists of lines
and 2-vertices where two lines meet complemented with exactly two additional
1-vertices from the field insertions. We label all vertices with distinct
integers. The lines are then naturally associated with pairs of integers. A
valid embedding on the lattice associates lattice sites with vertices such
that all lines map onto links, i.e. their index pairs refer to nearest
neighbor sites. The 1-vertices are at the sites $u$ and $v$ which can be
anywhere on the lattice.

We have found a representation for any embedded graph that we outline now.
This is by no means unique. Our representation will actually be quite
redundant by including extra information that will be found useful when we set
up an update scheme in the next subsection. With any $\Lambda$ we associate a
list $\ell$ which can be viewed as matrix $\ell_{i j}$. It has one row for
each vertex of the graph and there are five columns. In our convention the
first and second row are permanently associated with the 1-vertices at $u$ and
$v$. The remaining rows then deal with 2-vertices and the row-indices $i$ are
taken as the graph theoretic labels of the vertices. In the first column
$\ell_{i 1}$ we encode the lattice site where the vertex is embedded. To this
end we label the sites with integers $1, 2, \ldots, V$ in some order, with the
lattice volume given by
\begin{equation}
  V = \prod_{\mu} L_{\mu} .
\end{equation}
Rows $i$ of the list that are not in use for the given graph have $\ell_{i 1}
= 0$. The configuration $\Lambda$ in general holds many closed loops. The
columns $\ell_{i 2}$ and $\ell_{i 3}$ are filled such they allow to travel
around the loop containing a vertex $i$ by following the pointers
\begin{equation}
  i \rightarrow i' = \ell_{i 2} \rightarrow i'' = \ell_{i' 2} \rightarrow
  \ldots . \rightarrow i.
\end{equation}
The loop passing through $i$ can be traveled in two possible directions of
which one is given by column 2. By using column 3 one obtains the other
direction. Note that the loops of the O($N$) models are physically unoriented,
and we here encounter one of the redundancies mentioned before. The chain
between $u$ and $v$ is treated analogously with the exception that the
journeys are $1 \rightarrow i' = \ell_{1 2} \rightarrow i'' = \ell_{i' 2}
\rightarrow \ldots . \rightarrow 2$ and $2 \rightarrow i' = \ell_{2 3}
\rightarrow i'' = \ell_{i' 3} \rightarrow \ldots . \rightarrow 1$. From here
on we call this special sequence of vertices and lines the active
loop{\footnote{Remember it is a closed loop in the O($N$) sense yielding a
factor $N$.}} with the remaining ones being called passive. Column 4 just
holds a flag whose values distinguish vertices in passive loops (one) from
those in the active loop (zero). Finally column 5 is arranged such that the
following problem can be solved efficiently, i.e. without searching the whole
list: for a given lattice site{\footnote{By $x$ we designate both the
geometric location as well as the counting label given to it. }} $x$ find all
vertices (row-indices) embedded at this site for the present $\Lambda$. With
an additional entry-list $e (x)$ the problem is solved by following the
sequence
\begin{equation}
  x \rightarrow e (x) = i \rightarrow i' = \ell_{i 5} \rightarrow i'' =
  \ell_{i' 5} \rightarrow \ldots . \label{xchain}
\end{equation}
This chain ends when $\ell_{j 5} = 0$ is encountered. During the later update
a vertex can be removed from a graph. If it corresponds to row $i$, we set
$\ell_{i 1} = 0$ in such a case. Such a row can continue however to function
in ($\ref{xchain}$). As a consequence, in ($\ref{xchain}$) we can step through
such lines which are still needed even without holding a vertex. In addition
to $\ell$ and $e$ we also store the auxiliary field $d (x)$, although it could
be constructed from $\ell$.

\begin{figure}[htb]
  \begin{center}
    \epsfig{file=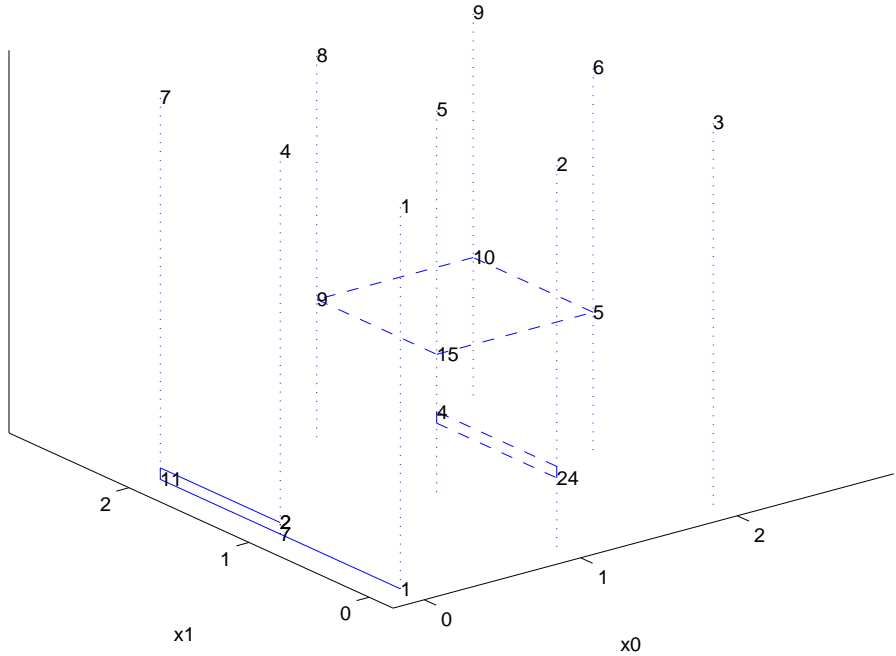,width=0.8\textwidth}
  \caption{A small loop configuration to illustrate its representation by a
  list.\label{fig1}}
\end{center}
\end{figure}

We now discuss as an example a $3 \times 3$ lattice taken from an O(3) run. It
is tiny to save space but it actually demonstrates most features. A loop
configuration is shown in figure \ref{fig1}. We have displaced the loops in
the third direction to disentangle overlapping loops. On the `roof' we show
the labeling of lattice sites, the other integers are vertex labels. The solid
line, connecting 1 with 2 (via 7 and 11) is the active loop and there are two
additional closed passive loops which overlap at site 5.

As already mentioned there is no hard bound on the number of rows/vertices
that are needed in $\ell$. In addition, as $\Lambda$ is updated (modified),
vertices are added and eliminated and thus rows of $\ell$ are freed (possibly
with the exception of the entry in column 5) and new ones are required. As the
available storage is finite, this requires some management which may seem
difficult at first sight. Luckily this problem can be handled rather easily.

As argued before the average total number of vertices will be of order $N
\times N_l$. As an extensive quantity its fluctuations are found to be only of
the order $\sqrt{N \times N_l}$. We thus found that reserving space for
\begin{equation}
  \{\ell_{i, j = 1, \ldots, 5} \}, i = 1, 2, \ldots, f_{\ell} \times N N_l
\end{equation}
lines with $f_{\ell}$ of order unity leads to completely negligible
probabilities to ever exhaust this space in any feasible simulation. These
storage requirements are quite similar to ordinary simulations. By observing
the fluctuations of $\sum_x d (x)$ one can easily demonstrate failure
probabilities like for example $10^{- 1000}$ which can be tolerated. We come
back to this in section 4.

For recycling list entries we keep another list with a reservoir of indices of
completely unused rows that are available for new vertices. There is a
subtlety here. As mentioned before a line not carrying a vertex anymore can
still be relevant as a `stepping stone' in (\ref{xchain}). Such rows we call
`unused' as opposed to `completely unused'. If we run out of completely unused
rows during a simulation we start a recycling routine which runs through the
whole list. In this process all $\ell_{i 5}$ where $\ell_{i 1} \not= 0$ and
the associated entries in $e$ are recomputed and all rows with $\ell_{i 1} =
0$ acquire the `completely unused' status. It turns out that the time spent
for list recycling is a negligible fraction of the total in practice. We end
this subsection by reproducing in table \ref{tab1} the lists $\ell$ and $e$
associated with the configuration in figure \ref{fig1}.

\begin{table}[htb]
\begin{center}
  \begin{tabular}{cccccc}
    1 & 1 & 7 & 0 & 0 & 0\\
    2 & 4 & 0 & 11 & 0 & 0\\
    3 & 0 & 21 & 1 & 0 & 24\\
    4 & 5 & 24 & 24 & 1 & 15\\
    5 & 6 & 10 & 15 & 1 & 8\\
    6 & 0 & 9 & 1 & 0 & 0\\
    7 & 4 & 11 & 1 & 0 & 16\\
    8 & 0 & 22 & 1 & 0 & 0\\
    9 & 8 & 15 & 10 & 1 & 0
  \end{tabular}{\hspace{2cm}}
  \begin{tabular}{cccccc}
    10 & 9 & 9 & 5 & 1 & 22\\
    11 & 7 & 2 & 7 & 0 & 17\\
    12 & 0 & 2 & 1 & 0 & 0\\
    15 & 5 & 5 & 9 & 1 & 6\\
    16 & 0 & 17 & 17 & 0 & 0\\
    17 & 0 & 7 & 7 & 0 & 0\\
    22 & 0 & 9 & 5 & 0 & 0\\
    24 & 2 & 4 & 4 & 1 & 0
  \end{tabular}
\end{center} 

    \begin{center}
      \begin{tabular}{ccccccccc}
        1 & 2 & 3 & 4 & 5 & 6 & 7 & 8 & 9\\
        12 & 3 & 0 & 7 & 4 & 5 & 11 & 9 & 10
      \end{tabular}
    \end{center}
 
  \caption{The list $\ell$ corresponding to the configuration in figure
  \ref{fig1} (upper two parts). It has been augmented by the leftmost
  (zeroth) column exhibiting the row indices. Completely unused lines have
  been omitted. The lower list is the entry table $e (x)$.\label{tab1} }
\end{table}

\subsection{Monte Carlo algorithm}

\begin{figure}[htb]
  \begin{center}
    \epsfig{file=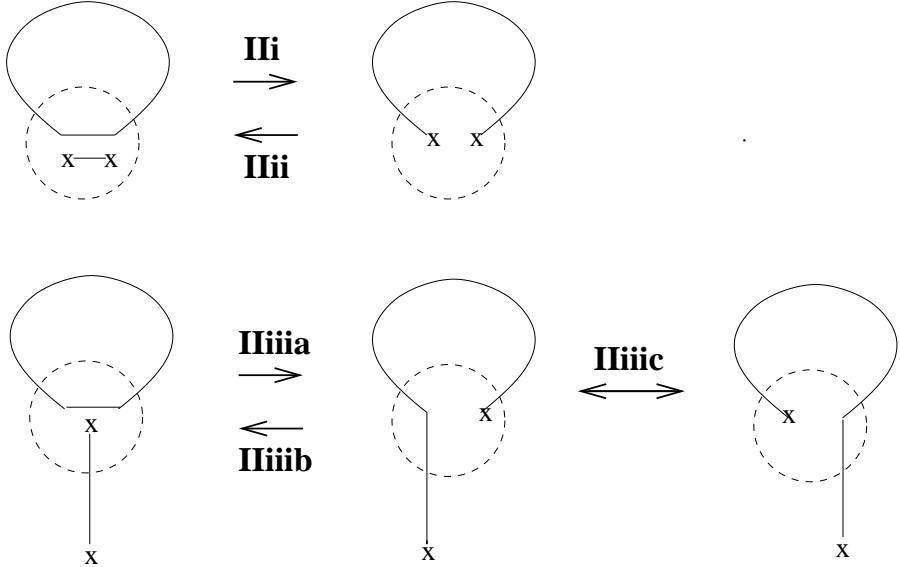,width=0.8\textwidth}
  \caption{Illustration for the re-routing moves. The dashed line encircles
  elements associated with one site. The x stand for the 1-vertices at $u, v$
  while a line both entering and leaving the dashed circle represents a
  2-vertex. There could be more `spectator' 2-vertices at the site which are
  not drawn for clarity.\label{algo}}
\end{center}
\end{figure}

We now propose an algorithm to simulate the O($N$) loop ensemble (\ref{LGav}).
We define a number of separate update steps such that each of them fulfills
detailed balance. They will then be iterated in some order as the final update
procedure. The moves are all Metropolis proposals for which we quote the ratio
$q$ which controls the
acceptance probability $\min (1, q)$ in each case. We need to introduce
the notion that the active loop between $u$ and $v$ is called trivial if it
contains no 2-vertex and also $u = v$ coincide.
\begin{enumerateromancap}
  \item Extension and retraction: We choose with equal probability between $2
  D + 1$ possible proposals to move $u$ by one lattice spacing with a
  concurrent adjustment of the active loop. In the first $2 D$ cases it is
  extended with $u$ moving to one of its neighbors $\tilde{u}$ with the amplitude
  ratio
  \begin{equation}
    q_{\tmop{ext}} = \frac{\beta}{N + d [ \text{$\tilde{u}$}]}  \frac{\rho (u
    - v)}{\rho ( \tilde{u} - v)} .
  \end{equation}
  In the last case $u$ is retracted by one link along the active loop with the
  ratio
  \begin{equation}
    q_{\tmop{ret}} = \frac{N + d [ \text{$u$}] - 2}{\beta}  \frac{\rho (u -
    v)}{\rho ( \tilde{u} - v)} .
  \end{equation}
  No move is made in the last case if the active loop is trivial.
  
  \item Re-route: We here want to change the O($N$) contraction or line
  connectivity structure at $u$. We have to distinguish a few cases. In the
  cases not covered below no move is made.
  \begin{enumerateroman}
    \item The active loop is trivial and $d (u) > 2$. In this case the we pick
    a 2-vertex at $u$ and replace it by the two 1-vertices, assigning $u$
    and $v$ randomly to the two lines.  The acceptance
    ratio is
    \begin{equation}
      q_{\tmop{rer}} = \frac{d [ \text{$u$}] - 2}{N} .
    \end{equation}
    Note that upon acceptance the active loop becomes non-trivial and the loop
    number $| \Lambda |$ is reduced by one.
    
    \item The active loop is not trivial but $u = v$ holds. In this case we
    make the move inverse to i with ratio
    $(q_{\tmop{rer}})^{- 1}$.
    
    \item We have $u \neq v$ and \ $d (u) > 2$. We pick with equal probability
    one of the lines connected to any of the 2-vertices at $u$ and propose to
    redirect it to the 1-vertex. The line previously connected to the latter
    is rewired to the newly created `opening' at the 2-vertex. For the
    acceptance decision we need to distinguish further sub-cases. In figure
    \ref{algo} we give a hopefully helpful illustration of the various moves.
    \begin{enumeratealpha}
      \item The chosen 2-vertex belongs to a passive loop. The latter then
      gets inserted into the active loop, $| \Lambda |$ is reduced by one, and
      the ratio is $1 / N$ in this case.
      
      \item The chosen 2-vertex belongs to the active loop, which
      self-intersects at $u$, and the chosen line leads towards the 1-vertex
      at $v$. In this case a new passive loop is detached and $| \Lambda |$
      goes up by one. The ratio is $N$.
      
      \item The chosen 2-vertex belongs to the active loop and the chosen line
      does not lead to the 1-vertex at $v$. In this case the active loop is
      just re-ordered and the ratio is one.
    \end{enumeratealpha}
  \end{enumerateroman}
  \item Kick: Do nothing unless the active loop is trivial. If this is the
  case we pick a random site $x$ and propose to move both $u = v$ to $x$. The
  active loop remains trivial. The acceptance amplitude ratio is
  \begin{equation}
    q_{\tmop{kick}} = \frac{N + d [ \text{$u$}] - 2}{N + d [ \text{$x$}]} .
  \end{equation}
\end{enumerateromancap}
\begin{figure}[htb]
  \begin{center}
    \epsfig{file=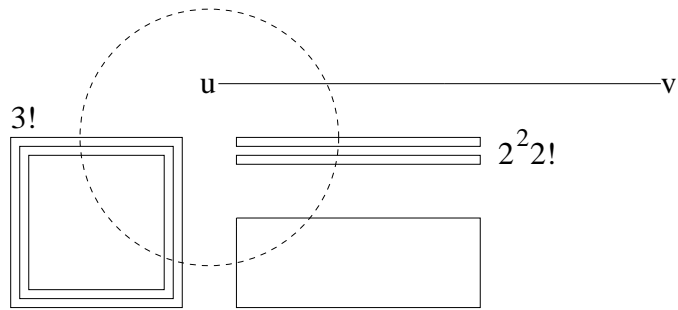,width=0.6\textwidth}
  \caption{A configuration which locally contributes a factor $3!\times 8$ to 
the symmetry factor $\mathcal{S}$. Vertices at $u$ are those inside the dashed circle.
Under moves of type IIiiia $\mathcal{S}$ can be reduced by factors 1, 3 or 4 in this case.
\label{symmfig}}
\end{center}
\end{figure}
The symmetry factor $\mathcal{S}[\Lambda]$ in (\ref{Wdef}) can only change during steps II,
because the active loop is distinguished and its changes do not influence the symmetry
of the graph $\Lambda$. If in II the symmetry changes, this is effectively 
taken into account by asymmetric proposal probabilities for the forward and backward process.
We explain this by an example given in figure~\ref{symmfig}.
In the figure contributions to the symmetry factor, where line permutations leave the
graph unchanged, are indicated. In step IIiii we first pick a line from one of the
2-vertices to then connect it to the 1-vertex at $u$. 
If lines participate in the symmetries several choices lead to the same
proposal $\Lambda\to\Lambda'$. More precisely the new graph is reached from
$\mathcal{S}[\Lambda]/\mathcal{S}[\Lambda']$ such choices. In this way the total
transition probability in IIiiia is
\begin{equation}
 p_{\rm a}=\frac{\mathcal{S}[\Lambda]/\mathcal{S}[\Lambda']}{d(u)-2} \min(1,1/N),
\end{equation}
while the reverse process IIiiib proceeds with
\begin{equation}
 p_{\rm b}=\frac{1}{d(u)-2} \min(1,N).
\end{equation}
In this way we have detailed balance with respect to (\ref{Wdef})
including $\mathcal{S}$. Analogous considerations apply to the other
moves in step II.

It should be easy to imagine now how the available information in the list
$\ell$ described before is useful during the update steps. For instance the
active/passive flag helps to discriminate between the sub-cases of IIiii based
on locally available information. It is also clear that precisely analogous
steps can be defined around $v$ instead of $u$ chosen above. After some brief
experiments we have arranged the updates in the following way
\begin{equation}
  1 \tmop{Iteration} : = (\Iota_u \Iota \Iota_u \Iota_v \Iota \Iota_v
  \tmop{III})^{N \times V / 2} . \label{Itdef}
\end{equation}
This was partly based on aesthetics and symmetry. \ For instance making only
$u$ steps or dropping III made little difference for our
observables{\footnote{If both is done, $v$ does not move anymore. Observables
where we average over translations still seemed to assume correct values.}}.
All Metropolis acceptance rates are well above 0.5 with the exception of the
extensions step in I. The latter is close to $(2 D)^{- 1}$. For simulations
with the modified action in section 5 it will rise to 1 however.

The mathematical proof of ergodicity is the usual one: with a nonzero
probability any configuration $\Lambda$ can be transformed to the trivial
empty one and then evolved to any other $\Lambda'$. It is also the empty
lattice from which we start all simulations. Another observation is the
following. A correct algorithm is also given with only the steps i and ii
contained in II, thus re-routing only for $u = v$. We have indeed confirmed
some correct results with such an update, but it is accompanied by severe
critical slowing down. We shall see that this is completely eliminated by the
additional steps{\footnote{In the zoological interpretation as a worm
algorithm {\cite{prokofev2001wacci}} step I corresponds to the normal
development of a worm. II deals with the asexual reproduction by detaching
parts of its body (IIiiib). Sadly, the O($N$) worm sometimes also devours its
offspring with probability $1 / N$ (IIiiia).}}.

Observables are accumulated after each $\Iota_u$ and $\Iota_v$ step during the
update. In most runs these contributions are stored separately for each
iteration, and later an off-line autocorrelation analysis
{\cite{Wolff:2003sm}} is carried out for these time series. We then arrive at
integrated autocorrelation times in units of iterations (\ref{Itdef}). During
the mostly local steps needed to update the list $\ell$ we sometimes have to
trace a closed loop, for example to adjust the active/passive flag. Because of
this it is not obvious at this point that the effort for one iteration scales
strictly proportional to the lattice size $V$. We shall come back to this
point in the next section.

\section{Numerical experiments with the standard action}

We have first made series of runs for the loop model representing the standard
lattice action discussed up to here. Most of these results can hence be
directly compared to numbers in the literature. In table \ref{tab2} we have
listed run parameters and the observed integrated autocorrelation times for a
few observables. For the A-series the $\beta$ values of {\cite{Wolff:1989hv}}
have been adopted together with lattice sizes to arrange for $m L \approx 8$
to hold. According to {\cite{Luscher:1983rk}} our massgap then differs from
the infinite volume one only at the level of $10^{- 4}$ which will be below
our errors. The weight $\rho$ in (\ref{LGav}) is chosen to roughly anticipate
the decay of the two point function. The {\tmem{relative}} error of the true
correlation (\ref{sscor}) is then constant or even shrinking as we explore the
exponential fall-off over a long range. We refer to the detailed discussion in
{\cite{Wolff:2008km}} which applies here unaltered. In this paper we use the
free massive lattice propagator to determine $\rho$,
\begin{equation}
  (- \Delta + \hat{M}^2) f (x) = \delta_{x, 0}  \hspace{1em} \Rightarrow
  \hspace{1em} \rho (x) = f (x) / f (0),
\end{equation}
where $\Delta$ is the standard nearest neighbor lattice Laplacian. Using the
fast Fourier transform on one lattice direction after another, its
construction costs negligible $\Omicron [D L^D \ln D]$ operations. Thus the
column for $\hat{M}$ completely determines $\rho (x)$.

\begin{table}[htb]
  \centering
  \begin{tabular}{|l|l|l|r|l|l|l|l|l|l|l|}
    \hline
    run & $N$ & $D$ & \multicolumn{1}{c|}{$L$} 
    & \multicolumn{1}{c|}{$\beta$}
    & \multicolumn{1}{c|}{$\hat{M}$}
    & $\tau_{\tmop{int}, K}$ &
    $\tau_{\tmop{int}, \chi}$ & $\tau_{\tmop{int}, m}$ & $\tau_{\tmop{int}, |
    \Lambda |}$ & CPU\\
    \hline\hline
    $\mathrm{A}_1$ & 3 & 2 & 56 & 1.4 & $8 L^{- 1}$ & 1.84(3) & 0.518(3) &
    0.678(6) & 1.46(2) & 1.50\\
    \hline
    $\mathrm{A}_2$ & 3 & 2 & 88 & 1.5 & $8 L^{- 1}$ & 1.74(3) & 0.515(3) &
    0.691(6) & 1.39(2) & 1.58\\
    \hline
    $\mathrm{A}_3$ & 3 & 2 & 152 & 1.6 & $8 L^{- 1}$ & 1.72(3) & 0.515(3) &
    0.698(6) & 1.38(2) & 1.75\\
    \hline
    $\mathrm{A}_4$ & 3 & 2 & 276 & 1.7 & $8 L^{- 1}$ & 1.71(3) & 0.514(3) &
    0.702(6) & 1.37(2) & 2.01\\
    \hline
    $\mathrm{A}_5$ & 3 & 2 & 518 & 1.8 & $8 L^{- 1}$ & 1.65(3) & 0.510(3) &
    0.689(6) & 1.34(2) & 2.56\\
    \hline
    $\mathrm{B}$ & 8 & 2 & 128 & 5.2 & $8 L^{- 1}$ & 1.94(3) & 0.506(3) & 0.571(4) &
    1.48(2) & 2.51\\
    \hline
    $\mathrm{C}_1$ & 1 & 2 & 128 & $\beta_c^{\tmop{ex}}$ & $\rho \equiv 1$ & 2.42(6) &
    0.927(14) & {\emdash} & 2.41(6) & 1.42\\
    \hline
    $\mathrm{C}_2$ & 1 & 2 & 256 & $\beta_c^{\tmop{ex}}$ & $\rho \equiv 1$ & 2.80(7) &
    1.005(16) & {\emdash} & 2.83(7) & 1.49\\
    \hline
    $\mathrm{D}_1$ & 3 & 3 & 32 &
    $\beta_c^{\text{{\cite{PhysRevE.72.016128}}}}$ & $\rho \equiv 1$ & 4.08(9)
    & 0.818(8) & {\emdash} & 3.02(6) & 1.50\\
    \hline
    $\mathrm{D}_2$ & 3 & 3 & 64 &
    $\beta_c^{\text{{\cite{PhysRevE.72.016128}}}}$ & $\rho \equiv 1$ &
    4.92(12) & 0.910(10) & {\emdash} & 3.63(8) & 1.83\\
    \hline
    $\mathrm{E}_1$ & 3 & 2 & 16 & 1.779 & $2 L^{- 1}$ & 0.824(8) & 0.536(4) &
    0.776(8) & 0.72(1) & 1.95\\
    \hline
    $\mathrm{E}_2$ & 3 & 2 & 32 & 1.779 & $2 L^{- 1}$ & 0.900(10) & 0.527(3) &
    0.842(9) & 0.79(1) & 2.21\\
    \hline
  \end{tabular}
  \caption{Run parameters and autocorrelation times. The statistics for each
  run consists of $10^6$ iterations. For A...D the geometry is symmetric
  $L_{\mu} \equiv L$, while in E it is elongated in `time', $L_0 = 6 L_1
  \equiv 6 L$. The critical value
  $\beta_c^{\text{{\cite{PhysRevE.72.016128}}}} = 0.693 002$ was taken from
  {\cite{PhysRevE.72.016128}}\label{tab2}.}
\end{table}

\begin{table}[htb]
  \centering
  \begin{tabular}{|l|l|l|l|l|r|}
    \hline
    run & \multicolumn{1}{c|}{$K$}
    & \multicolumn{1}{c|}{$\chi$}
    & \multicolumn{1}{c|}{$m^{- 1}$} 
    & $\langle \langle | \Lambda | \rangle
    \rangle_0 \times V^{- 1}$ & Ref.\\
    \hline\hline
    $\mathrm{A}_1$ & 0.78701(8) &  \ 78.75(16) &  \phantom{1}6.876(5) & 0.34590(4) &
    {\cite{Wolff:1989hv}}\\
    \hline
    $\mathrm{A}_2$ & 0.90246(6) &  175.38(41) & 11.053(8) & 0.34818(3) &
    {\cite{Wolff:1989hv}}\\
    \hline
    $\mathrm{A}_3$ & 1.01715(4) &  448.3(1.1) & 19.030(13) & 0.34604(2) &
    {\cite{Wolff:1989hv}}\\
    \hline
    $\mathrm{A}_4$ & 1.12920(2) & 1269.4(3.5) & 34.530(25) & 0.34244(1) &
    {\cite{Wolff:1989hv}}\\
    \hline
    $\mathrm{A}_5$ & 1.23829(1) & 3855(11) & 64.872(66) & 0.33927(1) &
    {\cite{Wolff:1989hv}}\\
    \hline
    B & 3.38678(6) & 431.54(80) & 18.096(9) & 0.88431(3) &
    {\cite{Wolff:1990dm}}\\
    \hline
    $\mathrm{C}_1$ & 0.31264(8) & 1.095(7)$L^{7 / 4}$ & {\emdash} & 0.13190(4) &
    {\cite{Wolff:2008km}}\\
    \hline
    $\mathrm{C}_2$ & 0.31217(5) & 1.104(8)$L^{7 / 4}$ & {\emdash} & 0.13216(3) &
    {\cite{Wolff:2008km}}\\
    \hline
    $\mathrm{D}_1$ & 0.23015(2) & 1.1055(19)$L^2$ & {\emdash} & 0.19294(1) &
    \\
    \hline
    $\mathrm{D}_2$ & 0.22910(1) & 1.0771(17)$L^2$ & {\emdash} & 0.193582(4) &
    \\
    \hline
    $\mathrm{E}_1$ & 1.2178(2) & 242.1(1.4) & 15.115(26) & 0.33892(11) &
    {\cite{Balog:2009np}}\\
    \hline
    $\mathrm{E}_2$ & 1.21662(9) & 632.5(3.1) & 25.115(37) & 0.33867(5) &
    {\cite{Balog:2009np}}\\
    \hline
  \end{tabular}
  \caption{Values of some physical observables defined in the text. In the
  last column we cite references, where data consistent with the ones here can
  be found\label{tab3}.}
\end{table}

In table \ref{tab3} the corresponding mean values and errors are compiled.
They refer to the mean bond occupation (\ref{kobs}), the susceptibility
(\ref{chiobs}), and the loop number appearing in (\ref{Wdef}). In addition we
have recorded the two point function
\begin{equation}
  G (t) = \langle \langle \rho (u - v) [\delta_{t, u_0 - v_0} + \delta_{t, u_1
  - v_1}] \rangle \rangle .
\end{equation}
In this formula, valid for the $D = 2$ symmetric geometry, we sum over both
directions and $\delta$ is taken $L$-periodic. In addition we average over
reflections. We have checked that these contributions are not completely, but
largely statistically independent.

\begin{figure}[htb]
  \begin{center}
    \epsfig{file=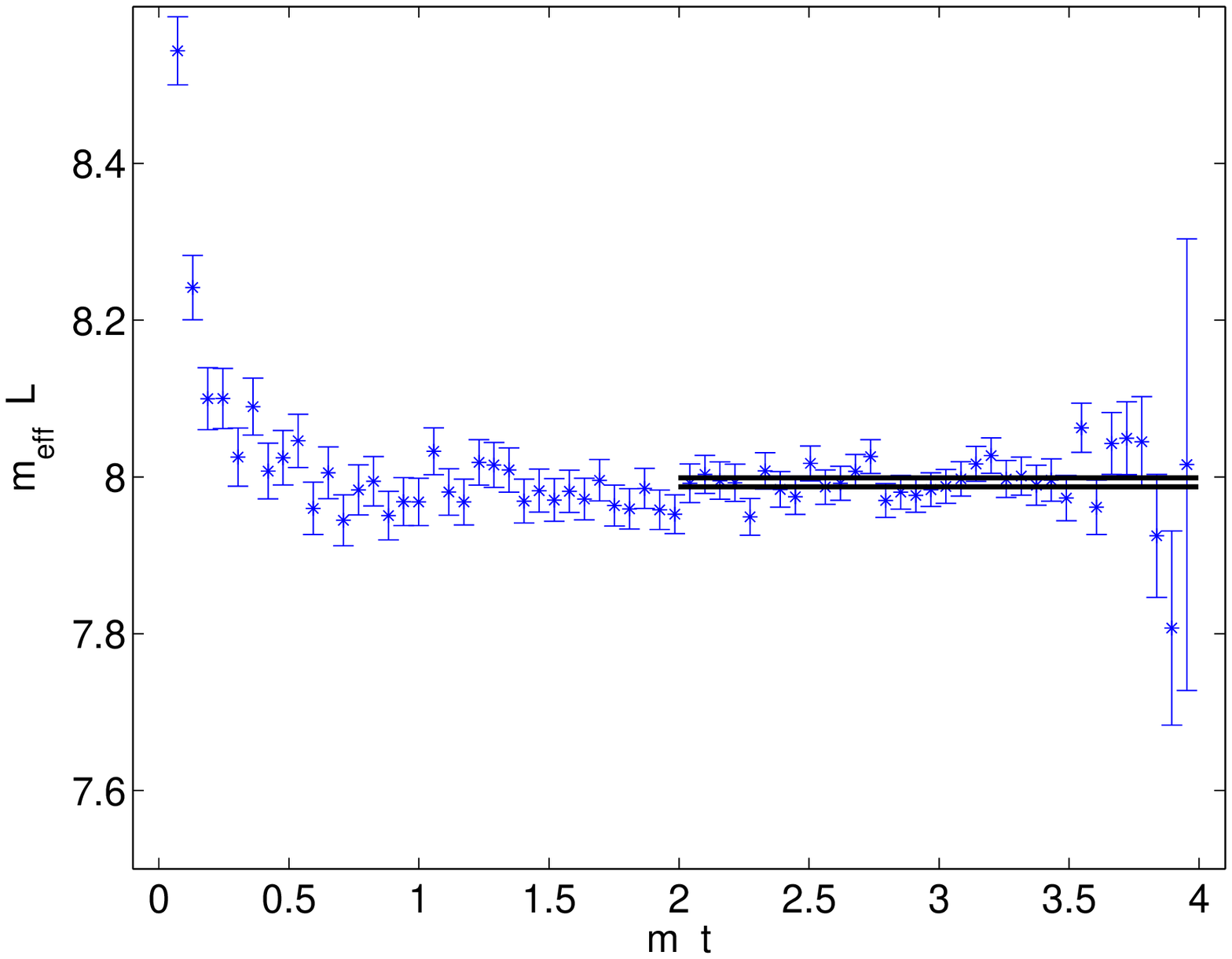,width=0.8\textwidth}
    \caption{Effective masses and fitted mass (band between the lines) from
    run $\mathrm{A}_4$ ($L = 276$). To avoid cluttering only every second
    $m_{\tmop{eff}}$ is shown.\label{fig2} }
\end{center}
\end{figure}

In figure \ref{fig2} the errorbars show effective masses derived from the
ratio of successive time-slice correlations by matching with $\cosh
(m_{\tmop{eff}} (t - L / 2))$. We see the expected long plateau and it seems
completely safe from excited states errors to finally extract the mass from a
fit, for which the horizontal line shows the range and the value in the form
of an $\pm 1 \sigma$ error band. For the fit we have minimized over the shown
range the function
\begin{equation}
  \Chi^2 = \sum_t \frac{[G (t) - c \cosh (m (t - L / 2))]^2}{\delta G (t)^2}
  \label{chisq}
\end{equation}
with respect to $c$ and $m$. More precisely, we first determine the error
$\delta G (t)$ by analyzing $G (t)$. As expected $\delta G (t)$/$G (t)$ hardly
grows with $t$. Then these errors are used to define via the minimization of
(\ref{chisq}) $m$ as a function of the primary correlation data. An error for
this derived observable is estimated as discussed in {\cite{Wolff:2003sm}}.
These are the values quoted for $m$ in table \ref{tab3}. Although the
effective masses fluctuate around the plateau (see figure \ref{fig2}), the $G
(t)$ cannot be expected to be completely uncorrelated at neighboring $t$
values. While (\ref{chisq}) is the so-called uncorrelated $\Chi^2$ we actually
see values between 0.1 and 0.4 per degree of freedom. This is not extremely
small, we collect much more independent information than is customary in
standard simulations. \ In fact the error of the fitted mass is between 2 (for
$\mathrm{A}_1$) and 4 (for $\mathrm{A}_5$) times smaller than that of the effective
mass at the beginning of the fit range alone. We have found that the mass
values and their errors change only within their errors, if we replace the
weight $\delta G^{- 2}$ in (\ref{chisq}) by one flat in $t$. Plots similar to
figure \ref{fig2} arise for all lattices where a mass is quoted. The data in
{\cite{Wolff:1989hv}} allow to roughly estimate which statistics was invested
for the errors quoted in units of steps per spin. We conclude that for the
estimation of the mass gap, the present method is quite competitive with the
reflection cluster algorithm {\cite{Wolff:1988uh}} with improved estimator
{\cite{Wolff:1989hv}}. This is not quite so for $\chi$, which could possibly
profit from a different choice of $\rho$.

In the series C we investigate the exactly solved two dimensional Ising model
by simply setting $N = 1$ in our loop code. The efficiency is in fact quite
similar to the simulations in {\cite{Wolff:2008km}} although the sampling of
the contraction structures in (\ref{LGav}) is an in principle unnecessary
complication as $N^{| \Lambda |}$ equals unity in this case, see the appendix
for further remarks. In the series D we take the O(3) model to three
dimensions and simulate at the critical point $\beta = 0.693 002$ determined
in {\cite{PhysRevE.72.016128}}. Beside the two runs quoted we have also
reproduced data from {\cite{Holm:1993zz}}. The C and D cases demonstrate the
very mild critical slowing down of our simulations at criticality where $L$ is
the only scale.

\begin{figure}[htb]
  \begin{center}
    \epsfig{file=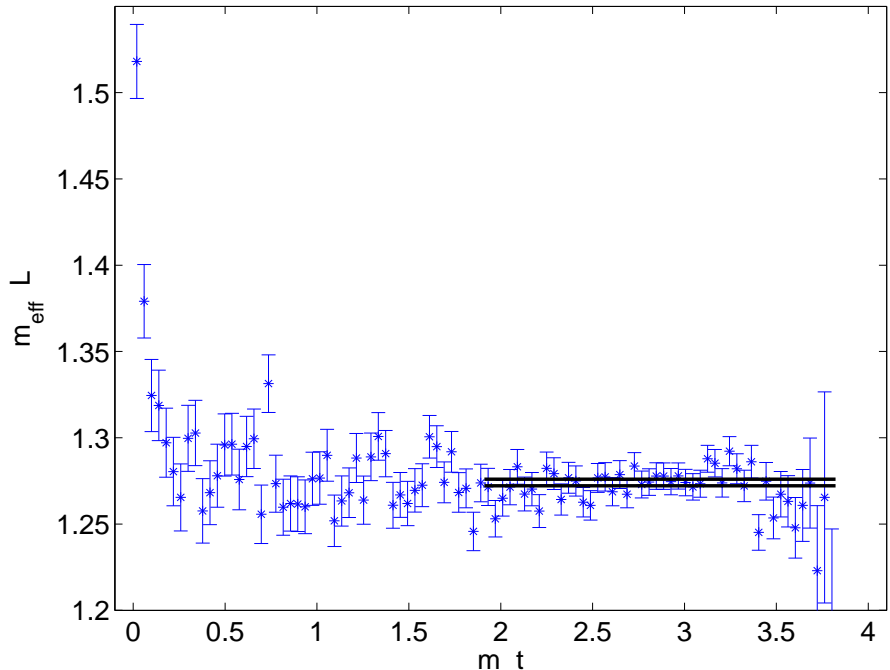,width=0.8\textwidth}
  \caption{Similar to figure \ref{fig2}, but for simulation $\mathrm{E}_2$ ($L_0
  = 192, L_1 = 32$).\label{fig4}}
\end{center}
\end{figure}

With E we turn to studies of the renormalized coupling
\begin{equation}
  \overline{g}^2 = m (L) L \hspace{1em} (N = 3) \label{lwwgbar}
\end{equation}
that has been introduced in {\cite{Luscher:1991wu}}. Here $m (L)$ is the mass
gap of the transfer matrix of spatial size $L$. In {\cite{Luscher:1991wu}} it
was argued that at least in the perturbative regime free boundary conditions
in the time direction help to isolate the first excited state from the rest of
the spectrum. The loop model could easily be modified to these boundary
conditions. We here prefer however to make the time direction long ($L_0 = 6
L$) to achieve the isolation. This is facilitated again by our small errors at
large separation. An advantage of this approach is that exact translation
invariance is kept for both directions. In figure \ref{fig4}, analogous to
\ref{fig2}, we demonstrate how the gap is extracted also in this case. Here we
took $\hat{M} > m$ and correspondingly `oversample' {\cite{Wolff:2008km}}
large distances with errors shrinking with growing $t$ (except very close to
$L_0 / 2$). Recently in {\cite{Balog:2009yj}} and {\cite{Balog:2009np}} a
thorough study of cutoff effects in the step scaling function for
(\ref{lwwgbar}) was made. In this context very accurate data were produced
including our lattices $\mathrm{E}_1, \mathrm{E}_2$ where we obtain
$\overline{g}^2 = 1.0586 (18)$ and $\overline{g}^2 = 1.2741 (19)$. In this
case a special estimator {\cite{Hasenbusch:1994rv}} is available in the spin
formulation such that our accuracy achieved here cannot really compete with
{\cite{Balog:2009np}}, but the more accurate values are consistent with ours.

We now make some general observations on the loop configurations observed in
our runs.

An exceedingly encouraging observation on our compiled autocorrelation times
is, that in units of iterations there seems to be almost no critical slowing
down for any of the series and quantities studied here. The typical bond
occupations $K$ resemble $E$, see (\ref{kobs}), and are functions of $\beta$
with only a weak dependence on $L_{\mu}$. They are larger in lower dimension.
The number of loops $| \Lambda |$ is strictly proportional to $V$ and also
similar to $E$. Both quantities grow with $N$ for similar correlation lengths
as one would expect.

We finally return to the question of the computational complexity of one
iteration. Here the last column `CPU' in table $\ref{tab2}$ is of interest. It
gives the execution time in $\mu \sec$ of one micro step, i.e. the time for
one iteration divided by $N V$. This is of course a highly non-universal
implementation and processor dependent quantity of not much interest. It is
quoted however, because its relative change between the runs may be of some
more interest. A constant value for CPU would suggest a scaling behavior like
for sweeps of local algorithms. The additional growth that we see within the
simulation series represents a small effective slowing down in CPU time units.
In the A series we really take a scaling limit. From $\mathrm{A}_1$ to $\mathrm{A}_5$
the correlation length changes by a factor 9.4. The extra CPU time factor is
1.7. The reason for this extra growth indeed comes from the `looping' steps as
discussed at the end of the last section. For a discussion of the true
asymptotic dynamical behavior, where this component will probably eventually
dominate, one could investigate the distribution and scaling behavior of the
perimeters of the loops in analogy to percolation cluster sizes. We however do
not try to determine dynamical exponents. We just conclude that the new type
of simulations is efficient enough to generate all data cited here in a few
hundred hours on a PC with a code that still allows for ample speedup. It is
not clear at the moment if a modified list parameterization and/or algorithm
could avoid this more than linear with the lattice volume growth of the CPU
time per iteration. Note that we have been discussing the continuum limit
throughout. For the thermodynamic limit ($V \rightarrow \infty$ at fixed
$\beta$) strict linearity is expected for the massive theory.

We close by a few remarks about our specific implementation that is reflected
in the quantity CPU. Our code is written in and running under
\tmtexttt{matlab}. Only the random numbers are imported from the C-code
{\cite{ranlux}} and we use luxury level two throughout. We used up-to-date
(2008) PCs with four cores that we employ for trivial parallelization thus
speeding up the time following from table \ref{tab2} by a factor four. We
hence always have four replica which allow for another reassuring consistency
check on the error determination by always monitoring the $Q$-values
{\cite{Wolff:2003sm}}. It would be clearly possible to speed up our runs by a
large factor by writing a dedicated C-code. In particular the data type
`pointer' seems ideal for the handling of loops by lists. Like for cluster
simulations, the present algorithm presumably does not lend itself very
naturally to a nontrivial parallelization.

\section{Simplified actions}

For the O($N$) model we have succeeded to simulate the untruncated strong
coupling expansion for the standard nearest neighbor lattice action. This
could be achieved because the weight needed to integrate out the spins
(\ref{Gdef}) is relatively simple in this case. We anticipate that in other
models that we shall want to treat similarly this may be more involved. It
would be easier to handle the graphs and to tabulate the required weights if
we could limit the overlapping of loops. For the O($N$) model this would
amount to constraining the sum in (\ref{Zsuv}) and in the formulae following
from it by
\begin{equation}
  k (l) = 0, 1, 2, \ldots, k_{\max} .
\end{equation}
In the simulations reported above the single link occupations $k (l)$ exhibit
a Poisson type distribution. Hence, for the mean values given, they never get
much beyond about 10 and a $k_{\max}$ of this size would have very little
effect. From here on we shall however investigate the most radical possibility
by setting $k_{\max} = 1$. Then we effectively replace (\ref{Zuv}) by
\begin{equation}
  \tilde{Z} (u, v) = \int \left[ \prod_z d \mu (s (z)) \right] s (u) \cdot s
  (v) \prod_{l = \langle x y \rangle} \left[ 1 + \tilde{\beta} s (x) \cdot s
  (y) \right]  \label{Zuv1}
\end{equation}
and use a tilde for quantities which refer to this action. It is clearly still
ultralocal and, if such a system becomes critical at all, we would expect to
be in the same universality class as before. The Boltzmann factor is strictly
positive for $| \tilde{\beta} | < 1$ only. In the Ising model at $N = 1$
the truncation $k_{\max} = 1$ is related to expanding in $\tilde{\beta} =
\tanh \beta$ instead of $\beta$. Then, with this identification, we compute
exactly the same correlations on every finite lattice.

A weight like (\ref{Zuv1}) is actually not new in the literature. In
{\cite{Domany:1981fg}} this action was put on a honeycomb lattice in two
dimensions. Since only three links meet at a site there, the strong coupling
graphs simplify as they cannot overlap or intersect, similarly to the Majorana
fermions in {\cite{Wolff:2008xa}}. In this way relations with certain discrete
models can be derived. This is elaborated in {\cite{Nienhuis:1982fx}} where
critical indices are derived for $- 2 \leqslant N \leqslant 2$. In both
publications it is conjectured that, although unusual, the Boltzmann weight in
(\ref{Zuv1}) should lead to the O($N$) universality class. We here check this
for the O(3) model in $D = 2$ (on our usual square lattice) by computing the
step scaling function (SSF) of {\cite{Luscher:1991wu}}.

It involves pairs of lattices
\begin{equation}
  \Sigma (2, u, L^{- 1}) = m (2 L) 2 L |_{m (L) L = u}
\end{equation}
where the side condition determines the value $\beta$ (or $\tilde{\beta}$) to
be used on both lattices. For each $u$ the SSF is expected to have a universal
continuum limit
\begin{equation}
  \sigma (2, u) \equiv \Sigma (2, u, 0) = \lim_{L \rightarrow \infty} \Sigma
  (2, u, L^{- 1})
\end{equation}
where $\sigma$ is equivalent to the Callan Symanzik beta function for the
renormalization scheme defined by the coupling (\ref{lwwgbar}). It is quite
amazing that by Bethe Ansatz techniques the continuum SSF $\sigma$ could be
computed exactly {\cite{Balog:2003yr}}. For our universality check we pick the
particularly popular point {\cite{Luscher:1991wu}}, {\cite{Balog:2009np}}
\begin{equation}
  \overline{u}_0 = 1.0595, \hspace{1em} \sigma (2, \overline{u}_0) = 1.261210.
\end{equation}
The exact result has been confirmed with extreme numerical precision in
{\cite{Balog:2009np}}.

A few short experiments immediately showed that $\tilde{\beta} > 1$ is
required to achieve any sizable correlation length in lattice units. Then the
weight in (\ref{Zuv1}) oscillates and presumably has a severe `sign problem'.
The loop partition function $\tilde{\mathcal{Z}}$ which is just as
(\ref{ZLam}) but omitting terms with any $\tilde{k} (l) > 1$ continues to be a
sum of positive terms, of course. It is also plausible that loops in
$\tilde{\Lambda}$ grow with rising $\tilde{\beta}$ and thus encode
correlations of growing range. The counterpart of (\ref{kobs}) now reads
\begin{equation}
  \tilde{K} = \frac{1}{N_l} \sum_l \langle \langle \tilde{k} (l) \rangle
  \rangle_0 = \frac{1}{N_l} \sum_{l = \langle x y \rangle} \left\langle
  \frac{\tilde{\beta} s (x) \cdot s (y)}{1 + \tilde{\beta} s (x) \cdot s (y)}
  \right\rangle .
\end{equation}
In the scaling region we shall find mean bond occupations not far from $1 /
2$. Due to the non-positive weight it is however not possible to draw any
immediate conclusions
on the nature of `typical' field configurations in the spin representation.
It also
seems hard to imagine to set up a bare perturbative expansion on the lattice
with this action. This is perhaps somewhat reminiscent of the constraint model in
{\cite{Hasenbusch:1995hu}} where the Boltzmann factor acts ferromagnetically
only in the form of a step function on the angle between neighboring spins. If
features of the long range physics are nevertheless described by renormalized
perturbation theory, this must be viewed more like an effective field theory
matched to these lattice models.

We now come to our concrete data\footnote{
Due to a programming error the data in this section had to be revised
in comparison to the previous arXiv versions.
} taken allowing $k (l) = 0, 1$ only. For the
SSF we have to first determine the values $\tilde{\beta}$ that yield
$\overline{g}^2$=$\overline{u}_0$ on a series of lattices. Such results are
listed in table $\ref{tab4}$. We have measured the derivative $\partial
\text{$\overline{g}^2 / \partial \tilde{\beta}$}$ and used it so solve the
problem of tuning $\tilde{\beta}$ (6th column). Our simulations are so close
to the target that further terms in a Taylor expansion and the errors of the
measured derivative play no r\^ole. We note that $\tilde{\beta}$ is larger
than the corresponding $\beta$ in $\mathrm{E}_{1,2}$
and that there are about four times fewer loops now. The effective mass plots were again inspected. They all look
qualitatively indistinguishable from figure \ref{fig4}. Again $\hat{M} = 2 /
L$ determined $\rho (u - v)$ for us.

\begin{table}[htb]
\centering
  \begin{tabular}{|r|r|c|c|c|c|c|}
    \hline
    \multicolumn{1}{|c|}{$L$} 
    & \multicolumn{1}{c|}{$\tilde{\beta}$} 
    & $\tilde{K}$ & $\langle | \tilde{\Lambda} |
    \rangle_0 \times V^{- 1}$ & $\overline{g}^2$ & $\tilde{\beta} (
    \overline{u}_0)$ & CPU\\
    \hline\hline
    6 &1.7851 &0.41112(11) &0.09098(5) &1.0622(9) &1.7904(18) &1.42 \\
    \hline
    8 &1.9107 &0.43875(9) &0.08468(5) &1.0612(10) &1.9144(21) &1.46 \\
    \hline
    12 &2.1114 &0.47897(6) &0.08125(4) &1.0622(10) &2.1185(26) &1.54 \\
    \hline
    16 &2.2841 &0.51044(5) &0.08060(3) &1.0636(10) &2.2955(28) &1.74 \\
    \hline
    24 &2.5742 &0.55930(3) &0.08073(2) &1.0590(10) &2.5724(36) &1.93 \\
    \hline
    50 &3.1055 &0.64402(2) &0.08165(1) &1.0597(11) &3.1063(32) &2.72 \\
   \hline
  \end{tabular}
  \caption{Lattice data with $k_{\max = 1}$ for the O(3) model in $D = 2$ with
  $L_0 = 6 L_1 \equiv 6 L$.  
  Each line represents $3 \times 10^6$ iterations.\label{tab4}}
\end{table}

In table \ref{tab5} we list the corresponding doubled lattices required for
the SSF.  The error quoted for
$\Sigma (2, \overline{u}_0, 1 / L)$ combines the statistical errors of both
the small and the doubled lattice.

\begin{table}[htb]
\centering
  \begin{tabular}{|r|r|c|c|c|c|c|}
    \hline
    \multicolumn{1}{|c|}{$L$} 
    & \multicolumn{1}{c|}{$\tilde{\beta}$} 
    & $\text{$\tilde{K}$}$ & $\langle | \tilde{\Lambda}
    | \rangle_0 \times V^{- 1}$ & $\overline{g}^2$ & $\Sigma (2,
    \overline{u}_0, 1 / L)$ & CPU\\
    \hline\hline
    12 &1.7904 &0.40987(10) &0.07897(5) &1.2300(17) &1.2300(22) &1.48 \\
    \hline
    16 &1.9144 &0.43832(7) &0.07880(4) &1.2304(17) &1.2304(23) &1.53 \\
    \hline
    24 &2.1185 &0.47978(5) &0.07913(3) &1.2424(17) &1.2424(23) &1.63 \\
    \hline
    32 &2.2955 &0.51219(4) &0.07966(2) &1.2448(18) &1.2448(23) &1.93 \\
    \hline
    48 &2.5724 &0.55882(3) &0.08045(2) &1.2547(18) &1.2547(25)  &2.15 \\
    \hline
    100 &3.1055 &0.64396(2) &0.08162(1) &1.2606(19) &1.2600(30) &3.34  \\
    \hline
  \end{tabular}
  \caption{The doubled lattices of table \ref{tab4}, $10^6$ iterations
  each.\label{tab5}}
\end{table}

\begin{figure}[htb]
  \begin{center}
    \epsfig{file=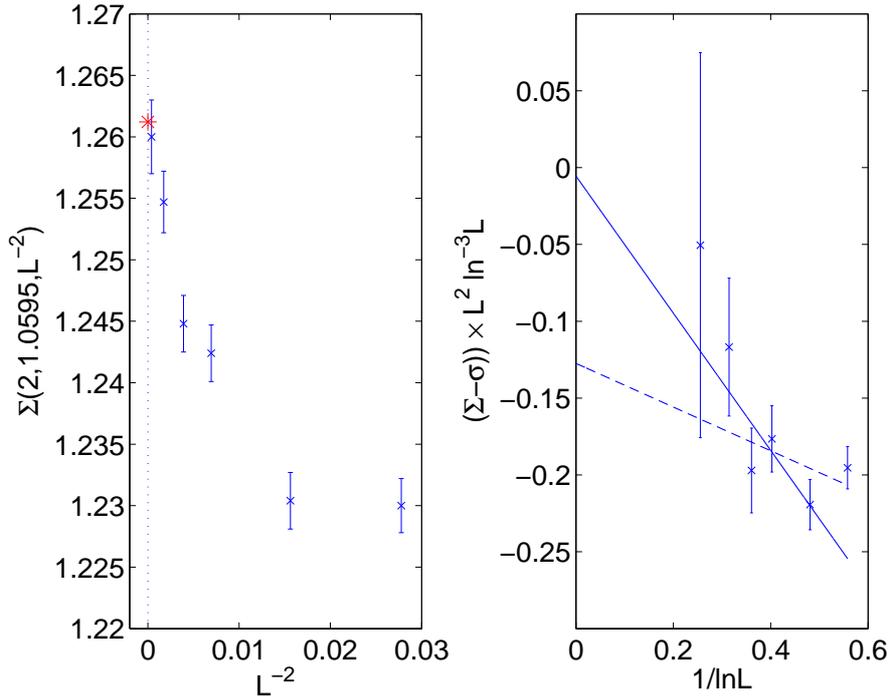,width=0.8\textwidth}
  \caption{Data for the step scaling function with the $k_{\max} = 1$ action.
  The star in the left plot is the known exact answer. The right plot is
  relevant for the theory of \ scaling violations \`a la Symanzik in
  {\cite{Balog:2009np}}.\label{fig5}}
\end{center}
\end{figure}

In figure \ref{fig5} we plot our SSF data. The left panel shows $\Sigma$
versus $L^{- 2}$ with the star at zero giving the exact continuum limit
{\cite{Balog:2003yr}}. It is very obvious that the data deriving from the
action in (\ref{Zuv1}) `know' about the universal value. In
{\cite{Balog:2009np}} a detailed investigation on the Symanzik effective field
theory for the lattice artifacts in the two dimensional O($N$) models was
carried out. The result is that the leading term for O(3) is expected to be of
the form $\ln^3 L / L^2$ with subleading contributions having smaller powers
of the logarithm. This has motivated us to plot $(\Sigma - \sigma) L^2 / \ln^3
L$ versus $\ln^{- 1} L$. The two lines are linear fits with 
acceptable $\Chi^2$ values. The dashed line differs
from the solid one by including/omitting the coarsest lattice pair. We see how
plotting matters: what is a short extrapolation in the left plot is an
uncomfortably long one in the right panel. But the artifacts for our action
are clearly compatible with the theoretical form. In addition it is reassuring
to note that any non-absurd extrapolation in the left plot would not miss the
continuum value by much. The finest lattice pair agrees with the continuum value
within our small statistical error.

\section{Conclusions and outlook}

We have successfully extended the method of simulating strong coupling graphs
instead of fields to the O($N$) models. They are in this way first
reformulated as a loop model which is then simulated. The resulting setup of
reformulation plus algorithm is, at least for the mass gap as an observable,
similarly efficient as the reflection cluster method. A variant of the lattice
Boltzmann factor, suggested by simplifying the loop ensemble, was simulated
and demonstrated to yield the same continuum step scaling function for a
particular value of the coupling. Hence universality is confirmed here in an
interesting case. A direct simulation as a spin model would presumably be
difficult due to sign oscillations of the Boltzmann weight.

The main goal of this work was not primarily to have another simulation method
for the O($N$) models. The main motivation was that the further extension of
this technique is possible in a more systematic way than for the cluster
method. After all, many systems possess a strong coupling expansion. The
O($N$) model is simply a very well studied case which has taught us important
lessons about the general approach. As a byproduct one could now however study
the O($N$) models for non-integer $N$ and perhaps one could also consider
taking the $N \rightarrow \infty$ limit (at fixed $\beta / N$) of the
algorithm similarly to the time continuum limit in {\cite{Beard:1996wj}}.
Also, as already indicated above, the study of the individual loop
distribution in the spirit of percolation theory could be of some interest.

A modest next step in the main line of the project will be to consider CP($N -
1$) spin models, for which no efficient cluster algorithm exists, at least for
the classical models (see however {\cite{Beard:2006mq}}). A more long distance
goal remains the simulation of lattice gauge theory as an equivalent surface
model.

{\noindent}\tmtextbf{Acknowledgments}: I would like to thank Martin
Hasenbusch, Erhard Seiler, Rainer Sommer and Peter Weisz for discussions and
help with the literature as well as Burkhard Bunk for advice with the
computers. Collaboration with Tomasz Korzec and Ferenc Niedermayer
was essential to correct the previous version of this eprint. 
Financial support of the DFG via SFB transregio 9 is acknowledged.

\appendix\section{The Ising limit $N = 1$ \label{appa}}

For $N = 1$ the O($N$) system becomes the Ising model. In this case the
function (\ref{Gdef}) is $G_1 (j) = \cosh (j)$ and hence
\begin{equation}
  c (n ; 1) = \frac{1}{(2 n) !}
\end{equation}
such that
\begin{equation}
  \prod_z c [d (z) / 2 ; 1] = \frac{1}{\mathcal{M}_0}
\end{equation}
may be used in (\ref{Wdef}). With no more dependence on $| \Lambda |$
\begin{equation}
  \mathcal{Z}= \sum_{\Lambda \in \mathcal{L}_2} \rho^{- 1} (u - v)
  \frac{\mathcal{M}[\Lambda]}{\mathcal{M}_0 [u, v ; k]} \left[ \prod_l
  \frac{\beta^{k (l)}}{k (l) !} \right]
\end{equation}
may be rewritten as a sum over $u, v, k$
\begin{equation}
  \mathcal{Z}= \sum_{u, v, k} \rho^{- 1} (u - v) \left[ \prod_l \frac{\beta^{k
  (l)}}{k (l) !} \right] \prod_x \delta_{d (x), \tmop{even}} .
  \label{ZuvIsing}
\end{equation}
which is the expected $\beta$ expansion for the Ising model. Note that $u, v$
enter via the relation (\ref{ddef}). \ While the form (\ref{ZLam}) contains
the sum over different contractions it is also correct with $N = 1$. All
contractions are equally likely in this case, the sum over them factorizes
off.

It is interesting to interpret our general update from the point of view of
the ensemble (\ref{ZuvIsing}). To obtain the effective transition probability
from step I we average over possible contractions at $u$ and sum over the
resulting ones at $\tilde{u}$. This leads \ (for $\rho \equiv 1$) to the
following update sequence
\begin{itemize}
  \item With probability $1 - 1 / (2 D + 1)$:
  \begin{itemize}
    \item Pick a link $l = \langle u \tilde{u} \rangle$ around $u$ in one of
    the $2 D$ directions with equal probability.
    
    \item Move $u \rightarrow \tilde{u}$ \ [and $k (l) \rightarrow k (l) +
    1$)] with probability\\
    $\min [1, \beta / (d ( \tilde{u}) + 1)]$.
  \end{itemize}
  \item With probability $1 / (2 D + 1)$:
  \begin{itemize}
    \item Pick a link $l = \langle u \tilde{u} \rangle$ around $u$ in one of
    the $2 D$ directions with probability $k (l) / (d (u) - 1)$. These
    probabilities add up to unity for the $2 D$ link directions if $u \not=
    v$ holds. For $u = v$ there is a probability $1 / (d (u) - 1)$ that no
    direction is picked. In this case no move is made.
    
    \item Otherwise move $u \rightarrow \tilde{u}$ \ [and $k (l) \rightarrow k
    (l) - 1$] with probability\\
    $\min [1, (d (u) - 1) / \beta]$
  \end{itemize}
\end{itemize}
The ratio in the third sub-item is the fraction of all possible pairings that
connects $u$ to a link in the given direction. It is not hard to see that
these implied transitions for $u, v, k$ obey detailed balance with respect to
(\ref{ZuvIsing}).

\end{document}